\documentclass[twocolumn,prl,amsmath,amssymb,showpacs]{revtex4}
\usepackage{graphicx}
\usepackage{dcolumn}
\usepackage{bm}
\usepackage{latexsym}
\usepackage{amstext}
\usepackage{amsxtra}
\newcommand{\bs}{\boldsymbol}

\begin{document}
\title{Realization of a magnetically guided atomic beam in the collisional regime}
\author{T. Lahaye, J. M. Vogels, K. G\"unter, Z. Wang, J. Dalibard,
and D. Gu\'ery-Odelin}
 \affiliation{Laboratoire Kastler Brossel$^{*}$, 24 rue Lhomond,
F-75231 Paris Cedex 05, France}
 \date{\today}

\begin{abstract}
We describe the realization of a magnetically guided beam of cold
rubidium atoms, with a flux of $7\times 10^9$ atoms/s, a
temperature of 400~$\mu$K and a mean velocity of 1~m/s. The rate
of elastic collisions within the beam is sufficient to ensure
thermalization. We show that the evaporation induced by a
radio-frequency wave leads to appreciable cooling and increase in
phase space density. We discuss the perspectives to reach the
quantum degenerate regime using evaporative cooling.
\end{abstract}


 \maketitle

The recent realization of slow atom sources using laser cooling
methods has constituted a major advance for many applications in
metrology and matter wave interferometry \cite{Interferometry}.
The possibility to complement laser cooling by forced evaporation
has led to even colder and denser atomic gases, and has culminated
with the observation of Bose-Einstein condensation (BEC)
\cite{BEC} and pulsed ``atom lasers" extracted from the condensate
\cite{atomlaser}. These advances in atom cooling allow in
principle to build a continuous and coherent source of matter
waves which would be the equivalent of a monochromatic laser. It
should lead to unprecedented performances in terms of focalization
or collimation, and provide a unique tool for future developments
in atom manipulation and precision measurements.

Two paths have been considered to reach the goal of a continuous
and coherent source of atoms (``cw atom laser''). The first
possibility \cite{ScienceK02} consists in using the ``standard"
condensation procedure to periodically replenish with new
condensates a gas sample held in an optical dipole trap. The
second possibility, that we are currently investigating, consists
in transposing the evaporative cooling method to a guided atomic
beam \cite{Mandonnet00}. As the beam progresses in a magnetic
guide \cite{mguide,cren}, the most energetic atoms are removed and
the remaining particles thermalize at a lower temperature and a
larger density, so that the beam can eventually reach the
degenerate regime. This method should in principle allow for a
large flux, since it consists in a parallel implementation of the
various steps to BEC. Instead of performing sequentially the
optical cooling and the evaporation sequences at the same place,
these occur simultaneously in different locations along the guided
beam. This gain in flux can constitute a major advantage for
applications.

To implement this method, the initial atomic beam needs to be
deeply in the collisional regime. The number $N_{\rm col }$ of
elastic collisions undergone by each atom as it travels along the
guide must be large compared to unity, to provide the
thermalization required for evaporative cooling. Such beams were
so far not available. In this Letter we report on the realization
of a rubidium atomic beam in the collisional regime, with $N_{\rm
col }\sim 8$. It propagates in a 4.5~m long magnetic guide with a
mean velocity of 1~m/s. We also demonstrate a first step of
evaporative cooling, and show that the measured increase in phase
space density is in good agreement with expectations. Finally we
discuss possible improvements of the current setup which should
allow to reach the degenerate regime.

Fig. 1 shows the layout of our experimental setup, composed of
three main parts lying in the $xy$ horizontal plane: the Zeeman
slower along the $y$ axis, the magneto-optical trap (MOT) and the
magnetic guide along the $x$ axis. The Zeeman slower
\cite{zeemanslower} consists in a 1.1~m long tapered solenoid in
which atoms emerging from an oven are decelerated by resonant
radiation pressure. It delivers a flux of $2\times
10^{11}$s$^{-1}$ $^{87}$Rb atoms, with a velocity of 20~m/s.

The slow atoms are captured in a 3-dimensional MOT formed by 3
pairs of counter-propagating beams. One pair of beam is aligned
with the vertical ($z$) axis, and the two other pairs are along
the directions $\bs e_x \pm \bs e_y$. Each trapping beam has a
circular profile, with a $1/e^2$ radius of 17~mm and a power of
15~mW. It is spatially filtered using a single mode optical fiber
and intensity stabilized to better than $1\%$. The MOT magnetic
gradients are $b'_x=0.6$~G/cm, $b'_y=5.4$~G/cm and $b'_z=-6$~G/cm.
The MOT is thus cigar-shaped, and its long axis coincides with the
magnetic guide direction. The capture rate of the MOT is $2\times
10^{10}$ atoms/s.

\begin{figure}[t]
\centerline{\includegraphics[width=80mm]{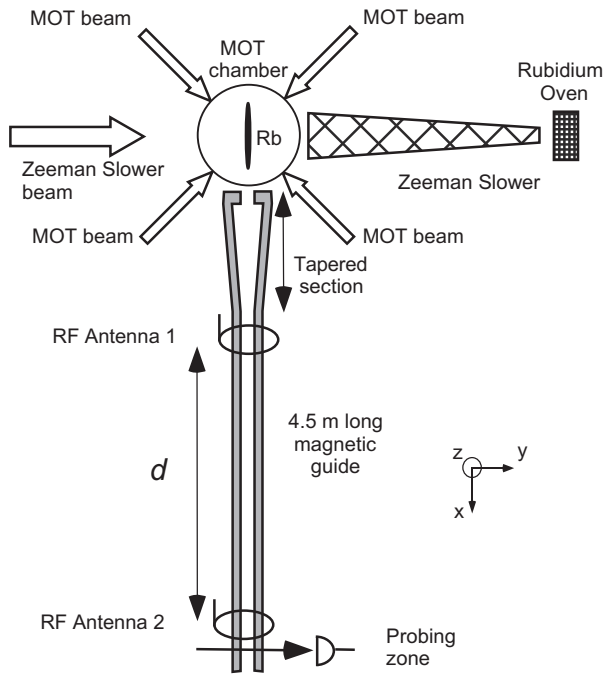}} \caption{Top
view of the setup. The atoms emerging from the oven are slowed
down by radiation pressure, captured in the magneto-optical trap,
and launched into the magnetic guide.} \label{fig1}
\end{figure}

The entrance of the magnetic guide is located 4.6 cm away from the
MOT center. The guide, which is placed inside the vacuum system,
consists of four parallel, water-cooled copper tubes, in which we
run a current $I$ up to $400$~A. Unless otherwise stated,
parameters and results given below correspond to $I=240$~A. The
entrance of the guide consists in a ring shaped metal part, that
is used for the recombination of the electric and cooling water
circuits \cite{cren}. The guide is held in place and shaped by
means of ceramic pieces. At the entrance of the guide the distance
between the centers of adjacent tubes is 14~mm, resulting in a
gradient of $b'_0=200$~G/cm. The first 40 cm of the magnetic guide
form a tapered section, obtained by reducing progressively the
distance between the tubes to 8~mm. This results in an increase of
the trapping gradient to $b'_1=$600~G/cm. The guide provides a
confinement of the atoms with a linear potential $\mu b'
(y^2+z^2)^{1/2}$, where $\mu$ is the magnetic moment of the atoms
\cite{Majorana}.

To optimize the transfer of the atoms from the MOT into the
magnetic guide, we operate in pulsed mode \cite{vogels}, with a
duration of $t_{\rm seq}=283$~ms per sequence. In the first 140~ms
we capture $3\times10^{9}$ atoms in the MOT. All beams are set at
the same frequency $\omega_L$, corresponding to a detuning
$\Delta=\omega_L-\omega_A=-3~\Gamma$, where $\Gamma/2\pi=5.9$~MHz
is the natural width of the $5P_{3/2}$ excited state and
$\omega_A$ the resonance frequency of a rubidium atom. The trapped
cloud is cigar-shaped, with a $1/\sqrt{e}$ radius of 1.4~mm and a
length of 35~mm.

The MOT magnetic field is then turned off and a 3~ms launching
phase occurs. The detunings of the beams in the $xy$ plane are
shifted by $\pm k_L v/\sqrt{2}$ with respect to the detuning
$\Delta$ of the vertical beams, where $k_L=\omega_L/c$. This forms
a moving molasses which sets the atoms in motion at velocity
$v=1.1$~m/s in the direction of the magnetic guide \cite{launch}.
The detuning $\Delta$ during the launching phase is ramped to
$-10~\Gamma$, which reduces the temperature of the gas to
$T_{0}=40 \,\mu$K in the frame moving at velocity $v$.

The final phase of the sequence consists in (i) optical pumping of
the moving atoms into the weak-field-seeking ground state
$|F=1,m_F=-1\rangle$, and (ii) magnetic pre-guiding of the atom
cloud. The pre-guiding is provided by a two-dimensional ($yz$)
quadrupole magnetic field $b'_2=65\,$G/cm generated by four
elongated coils (not shown in Fig. \ref{fig1}), located around the
MOT chamber. It prevents the atom cloud from expanding during its
flight towards the entrance of the magnetic guide. During the
pre-guiding phase we also apply at the MOT location a $56$~G bias
field along the $x$ axis to match the oscillation frequency in the
pre-guide with the cloud parameters. The overall efficiency of the
pumping+pre-guiding process is $60\%$ so that each packet loaded
into the guide contains $N_{\rm load}\sim 2\times 10^9$ atoms. We
measure the atom number and mean velocity by monitoring the
absorption of a resonant probe beam at the exit of the 4.5~m
magnetic guide. Losses due to collisions with residual gas are
negligible.

We have estimated numerically the influence of the rapid variation
of the magnetic gradient from $b'_2$ to $b'_0$ at the entrance in
the main guide by analyzing single particle trajectories. The
cloud is compressed to a final transverse radius $r_0=400\, \mu$m
and an energy per atom of $k_B\times 500\,\mu$K in the moving
frame. In the subsequent propagation into the tapered section of
the guide, the gradient increases to $b'_1$ and the bias field
decreases to 0 \cite{Majorana}. It leads to a further,
quasi-adiabatic, compression of the cloud and the energy per atom
raises to $k_B\times 1400$~$\mu$K. Assuming that thermalization
occurs via elastic collisions, this corresponds to a temperature
of $T_1=400\,\mu$K shared as $3k_BT_1/2$ for the kinetic energy in
the moving frame and $2k_BT_1$ for the potential energy of the
transverse linear confinement. This estimate for $T_1$ is in good
agreement with our measurement as we shall see below.

\begin{figure}[t]
\centerline{\includegraphics[width=80mm]{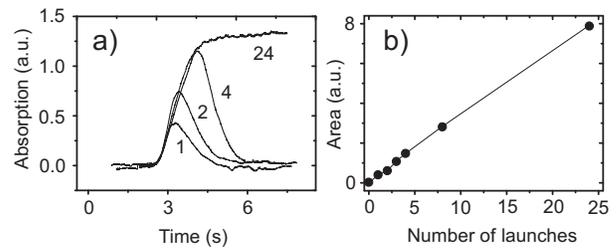}}
 \caption{a) Absorption signal measured at the end of the guide for 1, 2, 4 and 24 launches. b) Number of atoms
(area of the absorption signal) as a function of the number of
launches. The time origin corresponds to the launching of the
first packet.} \label{fig2}
\end{figure}

The next capture+launch sequence starts 140~ms after the beginning
of the pre-guiding phase, so that the atom packet has moved $\sim
14$~cm away from the MOT center and is now well inside the main
guide. A remarkable feature of this protocol is that the light
scattered from the MOT during the preparation of a packet does not
produce detectable losses on the previous packet. Fig.~\ref{fig2}
shows the signal detected for different numbers of launches. One
can check that the area of the signal scales linearly with this
number.

To generate a continuous beam we repeat the sequence with a
cycling rate of $t_{\rm seq}^{-1}=3.5$~Hz. For $T=400\,\mu$K the
packets overlap after a propagation of 60~cm. The flux is
$\phi=N_{\rm load} t_{\rm seq}^{-1}= 7\,(\pm 2) \times 10^{9}$
atoms/s.

We now turn to the characterization of the beam and the
demonstration that the collisional regime has indeed been reached.
The main tools for this study are radio-frequency (r.f.) antennas
located along the guide, which generate an oscillating magnetic
field at frequency $\nu$. An antenna flips the magnetic moment of
the atoms passing at a distance $r_{\rm e}=h\nu/(\mu b')$ from the
$x$ axis (r.f. evaporation \cite{rfevaporation}). For an atom with
transverse energy $E$ and angular momentum $L$ along the $x$ axis,
the evaporation criterion $r=r_{\rm e}$ is fulfilled at some
points of the atom trajectory if
\begin{equation}
 E\; \geq \; \frac{L^2}{2mr_{\rm e}^2}+\mu b' r_{\rm e}\ ,
 \label{evapcond}
\end{equation}
where $m$ is the atom mass. The range of an antenna is $\sim
20$~cm, so that it constitutes a local probe of the $(E,L)$
distribution at the scale of the whole guide.

\begin{figure}[t]
\centerline{\includegraphics[width=85mm]{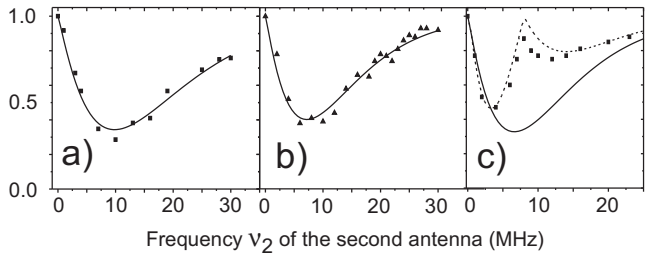}} \caption{
Experiment with two antennas separated by a distance $d$:
remaining fraction after the second antenna, with the first
antenna set at a fixed value $\nu_1$. a) Antenna 1 off, flux
$\phi$, $b'_1=600$~G/cm and $d=3$~m.  Using (\ref{fitfunction}) as
a fit function (solid line), we find $T_1=384\, (\pm\, 13)\,\mu$K.
b) $\nu_1=24$~MHz with the same conditions as for a). The fit
gives $T_2=281 \,(\pm8)\;\mu$K indicating a significant cooling.
c) $\nu_1=8$~MHz and reduced number of collisions between
antennas: flux $\phi/3$, $b'_1=400$~G/cm, $d=0.85$~m. The dotted
line is the prediction for the collisionless regime. The solid
line is the result expected if thermalization had occurred.}
\label{fig3}
\end{figure}

The temperature in the guide is deduced from the variation of the
fraction $f(\nu)$ of remaining atoms after the antenna. For
$r_{\rm e}$ small compared to the thermal size $r_{\rm
th}=2k_BT/(\mu b')$, only a very small fraction of atoms has a
sufficiently low angular momentum to be evaporated, as a
consequence $f(\nu\rightarrow 0)=1$. In the opposite limit $r_{\rm
e}\gg r_{\rm th}$, the evaporated fraction is negligible because
of the exponential decay of the energy distribution, thus
$f(\infty)=1$. We have used a Monte-Carlo determination of
$f(\nu)$, that is accurately fitted by the analytical form:
 \begin{equation}
f(\nu)=1-A\,\eta^{B}\,e^{-C\eta} \quad \mbox{ with }\quad
\eta=h\nu/(k_BT)
 \label{fitfunction}
 \end{equation}
with $A=1.65$, $B=1.13$ and $C=0.92$. A minimum value $f^*\simeq
0.33$ is expected for $ h\nu^*/(k_BT) \simeq 1.25\,$.

A typical measurement of $f(\nu)$ is given in Fig. 3a, together
with a fit using (\ref{fitfunction}). The transverse temperature
corresponding to the best fit is $384\, (\pm\, 13)\,\mu$K, in good
agreement with the expected temperature given above. For the fit
we allow the coefficient $A$ in (\ref{fitfunction}) to vary, to
account for a possible uncomplete efficiency of the evaporation
process. For the data of Fig. \ref{fig3}a, the efficiency is
better than 90\%.

The thermal equilibrium assumption used above is valid only if a
sufficient number of elastic collisions (at least a few per atom)
has occurred in the guide. The good agreement between the shapes
of the experimental and predicted functions $f(\nu)$ is a first
indication that the collisional regime has indeed been reached. To
investigate this point further, we now turn to a two-antenna
experiment (Fig.~\ref{fig3}b). These antennas are located
respectively $d_1=1$~m and $d_2=4$~m after the entrance of the
guide. The frequency $\nu_1$ of the first antenna is fixed at
$\nu_1=24$ MHz, which corresponds to $\eta_1\equiv
h\nu_1/(k_BT_1)=3$, where $T_1$ is the temperature deduced from
the analysis of Fig.~\ref{fig3}a. This antenna evaporates the
atoms fulfilling (\ref{evapcond}), thus producing a non thermal
($E,L)$ distribution. We measure the function $f(\nu_2)$ with the
second antenna, and check whether a thermal distribution is
recovered during the propagation from $d_1$ to $d_2$. The result
for $f(\nu_2)$ in Fig.~\ref{fig3}b shows clearly that this is
indeed the case. We infer from this set of data a temperature
$T_2=281 \,(\pm8)\;\mu$K. The fact that $T_2<T_1$ is a clear
demonstration of evaporative cooling of the beam by the first
antenna.

\begin{figure}[t]
\centerline{\includegraphics[width=80mm]{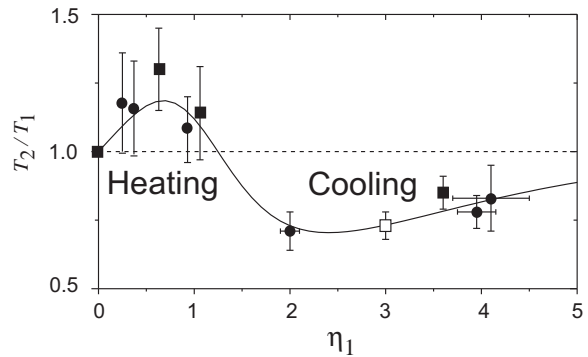}}
 \caption{Ratio $T_2/T_1$ between the
temperature $T_2$ of the beam when the first antenna is set at
$\nu_1$ and  the temperature $T_1$ of the beam in absence of
evaporation, as a function of the dimensionless parameter
$\eta_1=h\nu_1/(k_BT_1)$. The solid line is the theoretical
prediction for a linear potential (no adjustable parameter). The
open square corresponds to the data of Fig. \ref{fig3}b. The
squares (resp. circles) have been obtained with $I=240$~A (resp.
$400$~A).}
 \label{fig4}
\end{figure}

We have investigated the thermalization process as a function of
the evaporation frequency $\nu_1$ (Fig.~\ref{fig4}). For $\eta_1
\lesssim 1.25$ the evaporated atoms have less energy than average,
leading to heating ($T_2>T_1$). Cooling is observed in the
opposite case $\eta_1 \gtrsim 1.25$. The continuous line in
Fig.~\ref{fig4} is the theoretical prediction assuming full
thermalization between $d_1$ and $d_2$. It is in excellent
agreement with the experimental results. The phase space density
at the center of the beam scales as $\phi/T^{7/2}$. When we choose
$\eta_1=3$ (Fig. \ref{fig3}b), we measure a decrease of the flux
by $36\%$ and a decrease of temperature of $27\%$. This
corresponds to an \emph{increase} in phase space by a factor
$1.92\,(\pm \,0.15)$, to be compared with the expected gain of
1.90. It shows that this transverse evaporation mechanism is
indeed well suited for our final goal, i.e. bringing the beam in
the degenerate regime.

We have also investigated the transition between the collisional
and collisionless regimes. For this study we have reduced the
collision rate by decreasing the flux to $\phi/3$ and by operating
the magnetic guide at a lower current (160~A). The measured
temperature is in this case $T_1=230\,(\pm\,6)\;\mu$K. To minimize
the thermalization process between the antennas, we have also
reduced the distance $d_2-d_1$ to 0.85~m. The result for
$f(\nu_2)$ is shown in Fig.~\ref{fig3}c for the particular choice
$\nu_1=8\,$MHz. The dashed line gives the result expected in the
collisionless regime. In this case the second antenna does not
induce extra loss when $\nu_2=\nu_1$ since no modification of the
$(E,L)$ distribution occurs between the two antennas. The measured
$f(\nu_2)$ is in good agreement with this prediction. For
comparison we also indicate in Fig.~\ref{fig3}c the expected
signal in case of a complete thermalization between the antennas.
It clearly does not agree with our data, contrary to the large
flux case (Fig.~\ref{fig3}b).

To summarize, our guided beam is deeply in the collisional regime
when operated at the maximum flux. The collision rate in the
linear guide is $\gamma_c=n_0\, \sigma\, \Delta
v/\sqrt{\pi}=2\,$s$^{-1}$, where $n_0=2.8\times
10^{10}\,$cm$^{-3}$ is the central density, $\sigma=6.8\times
10^{-12}\,$cm$^2$ is the s-wave elastic cross-section at low
temperature for spin polarized $^{87}$Rb atoms and $\Delta
v=\sqrt{k_BT/ m}=19\,$cm/s. This corresponds to $N_{\rm col}=8$
collisions per atom along the whole guide. For the thermalization
experiment of Fig. \ref{fig3}b and taking into account the
reduction of flux by the first antenna, we estimate that each atom
undergoes 3 collisions between the two antennas. This is known to
be the minimum value to ensure thermalization (see e.g.
\cite{timecoll}).

Finally we briefly discuss the possibility to use this device for
producing a coherent continuous atomic wave. The current phase
space density at the center of the beam is $2\times 10^{-8}$. To
reach quantum degeneracy, we plan to use several antennas with
decreasing frequencies. Operating all antennas at the same $\eta$,
the cooling distance is minimized for $\eta\sim\,$4. A total of 40
antennas (each providing a phase space increase of 1.63) is
needed. The expected output flux is $2\times 10^{-4}\phi$ for a
cooling distance of $\sim 100\, v/\gamma_c$. To match this
distance with the length of our guide, we need to increase the
initial collision rate $\gamma_c$ by one order of magnitude. A
fraction of this gain can be obtained by slowing down the beam by
gravity, using a bent or tilted guide \cite{ldea}. In addition we
plan to transpose to our elongated MOT the ``dark spot" technique,
which is known to lead to a significant gain in terms of spatial
density of the laser cooled atomic source \cite{Darkspot}. A
continuous coherent atomic beam will be a very useful tool not
only for the interferometric and metrology applications mentioned
above, but also for studies of fundamental issues, such as the
existence of superfluidity in quasi one-dimensional systems.
Another exciting challenge will be the miniaturization of the
setup, along the line of what has been achieved recently with
``atom chips'' for the production of standard BEC \cite{chips}.

We acknowledge stimulating discussion with the ENS laser cooling
group. We thank A. Senger for his participation to the data
acquisition. K. G\"unter acknowledges support from the Marie Curie
Host Fellowship ``QPAF'' HPMT-CT-2000-00102. J. M. Vogels
acknowledges support from the Research Training Network ``Cold
Quantum Gases'' HPRN-CT-2000-00125. This work was partially
supported by the CNRS, the R\'egion Ile-de-France, the Coll\`{e}ge
de France, the Ecole Normale Sup\'erieure, the Bureau National de
la M\'etrologie, the University of Paris VI and the D\'el\'egation
G\'en\'erale de l'Armement.


\begin{thebibliography}{99}
\bibitem{Majorana}
Strictly speaking, the magnetic field at the center of the guide
does not vanish thanks to the $x$ component of the Earth magnetic
field, which  prevents Majorana spin flips.
\end{thebibliography}
\end{document}